# Spanners for Geometric Intersection Graphs

Martin Fürer  Shiva Prasad Kasiviswanathan

Computer Science and Engineering, Pennsylvania State University
e-mail: {furer,kasivisw}@cse.psu.edu

## Abstract

Efficient algorithms are presented for constructing spanners in geometric intersection graphs. For a unit ball graph in $\mathbb{R}^k$, a $(1+\epsilon)$-spanner with $O(n\epsilon^{-k+1})$ edges is obtained using efficient partitioning of the space into hypercubes and solving bichromatic closest pair problems. The spanner construction has almost equivalent complexity to the construction of Euclidean minimum spanning trees. The results are extended to arbitrary ball graphs with a sub-quadratic running time.

For unit ball graphs, the spanners have a small separator decomposition which can be used to obtain efficient algorithms for approximating proximity problems like diameter and distance queries. The results on compressed quadtrees, geometric graph separators, and diameter approximation might be of independent interest.

## 1 Introduction

Let $G = (V, E)$ be a weighted graph, and let $d_G(u, v)$ be the length of shortest path between vertices $u$ and $v$ in $G$. For any fixed $\epsilon > 0$, a stretch $(1 + \epsilon)$-spanner of $G$ is a subgraph $G'$ such that for all pairs of vertices $u$ and $v$, $d_{G'}(u, v)/d_G(u, v) \leq (1 + \epsilon)$. Spanners are important structures because they provide a way for approximating the graph in an economical way. Spanner constructions have been widely investigated for general graphs, and complete Euclidean graphs, also with additional properties like weight, diameter, degree [ACIM99, CK95, ADM+95].

We present a new method for producing spanners of geometric graphs based on a hierarchical decomposition of the plane into tiles of various sizes. Our constructions are also more general, as they are not restricted to complete Euclidean graphs, but extend to geometric unit disk and arbitrary disk graphs, as well as their higher dimensional versions, the ball graphs. In all cases, edge lengths are given by Euclidean distances, but not all edges have to be present in our graphs. The difficulty in constructing a spanner for the disk graph metric when compared to the metric induced by a complete Euclidean graph is that two points that are close in space are not necessarily close under the graph metric.

Intersection graphs are graphs whose vertices are represented by sets such that two vertices are adjacent if and only if the corresponding sets have a non-empty intersection. A disk graph is an intersection graph of disks in the plane. We consider weighted disk graphs where the weight of an edge is the Euclidean distance between centers. Such graphs have been used widely to model the communication between objects in VLSI [MC80] and recently in the context of wireless networks [KMR01, Li03]. In the wireless networks they model the fact that two wireless nodes can directly communicate with each other only if they are within certain distance. Many restricted versions of disk graphs like unit disk graphs, quasi unit disk graphs, $\lambda$-precision unit disk graphs have also been investigated as models for wireless networks [Li03, KZ03, HMR+98].

Spanners are important for disk graphs because restricting the size of a network reduces the amount of routing information. One of the requirements of topology control is to produce a low stretch spanner [Raj02, Li03]. Li *et al.* [LWW01] showed that deleting certain edges from the Yao-graph (introduced in [Yao82]) results in a bounded stretch and bounded degree spanner for the special case where all disks have unit radius. For solving proximity problems Gao and Zhang [GZ05] gave a construction of a c-well-separated pair decomposition (introduced in [CK95]) for unit disk graphs with $O(n \log n)$ pairs, for any constant $c \geq 1$. They also show that in $\mathbb{R}^k$ at least $O(n^{2-2/k})$ pairs are needed. However, for general disk graphs no sub-quadratic well-separated pair-decomposition is possible.

## 2 Preliminaries

Let $\mathcal{P}$ be a set of points in $\mathbb{R}^k$ for any fixed dimension $k$. Let $\mathcal{D}$ be a set of $n$ balls such that (i) $D_p$ is centered at $p \in \mathcal{P}$, and (ii) $D_p$ has radius of $r_p$. Balls $D_p$ and $D_q$ intersect if $d(p,q) \leq (r_p + r_q)$, where $d(.,.)$ denotes the Euclidean metric. The disk graph $G$ is a weighted graph where an edge $e = (p,q)$ with weight $d(p,q)$ exists if $D_p$ and $D_q$ intersect. Let $d_G$ denote the metric induced by the connected graph $G$ on its nodes by shortest paths. We re-scale the balls such that the largest radius equals one. The global stretch factor of $\mathcal{D}$ is then defined as $\rho(\mathcal{D}) = 1/\min\{r_p : D_p \in \mathcal{D}\}$.

In the plane our algorithms use a variant of quadtrees. For a node $t$, denote by $P(t)$ the parent of $t$ in the tree. A point $(x,y)$ is *contained* in a node $t$ representing a square with center $(x_t, y_t)$ and length $l_t$ in the quadtree iff $x_t - l_t/2 \leq x < x_t + l_t/2$ and $y_t - l_t/2 \leq y < y_t + l_t/2$. For a set of squares $T$ in the quadtree a point is contained in $T$ iff there exists $t \in T$, such that point is contained in $t$.

To avoid ambiguities, throughout the paper we refer to the vertices of a graph as *vertices* and vertices of a tree as *nodes*. We assume w.l.o.g. that $\epsilon^{-1}$ is a power of 2. Floors and ceilings are omitted throughout the paper, unless needed. Note that starting with a fixed $\epsilon$ we get a $(1 + c\epsilon)$ spanner for a fixed constant $c$. For simplicity we describe all the algorithms for $k = 2$ and then state the generalizations to higher $k$. We use the notation $\tilde{O}(f) \equiv O(f \operatorname{poly} \log f)$.

### 2.1 Modified Yao-graph and Our Results

We use a modified version of the Yao-graph construction. Let $\mathcal{C}(p) = \{c_1(p), \ldots, c_{\epsilon^{-1}}(p)\}$ be a collection of $\epsilon^{-1}$ cones such that (i) each cone has apex at $p \in \mathcal{P}$, (ii) each cone has an opening angle of $2\pi\epsilon$, and (iii) the union of these cones covers $\mathbb{R}^2$. We define a modified Yao-graph $\overrightarrow{Y}$ in the following way. The vertices of $\overrightarrow{Y}$ are the points of $\mathcal{P}$. For each $p \in \mathcal{P}$ and $1 \leq i \leq \epsilon^{-1}$ add a directed edge from $p$ to point $q$ contained in $c_i(p)$ if: (i) $q$ is closest to $p$ in $c_i(p)$, (ii) $r_q \geq r_p$, and (iii) the edge $(p,q)$ exists in $G$. We use $Y$ to denote the undirected counterpart of $\overrightarrow{Y}$.

**Lemma 1.** *Let $(u,v)$ be an edge in the disk graph $G$. Then there exists a path in the spanner $Y$ such that $d_Y(u,v) \leq (1+\epsilon)d(u,v)$.*

**Proof.** Let $r_u \leq r_v$. Consider the cone $c_i(u)$ containing $v$. Let $(u,w)$ be the edge added in $\overrightarrow{Y}$ from $u$ to $w$ contained in $c_i(u)$. The distance $d(w,v) \leq d(u,v) - (1 - O(\epsilon))d(u,w)$. Since $r_w \geq r_u$, we know that there exists an edge $(w,v)$ in $G$. The proof can be completed by using induction over distances. ❑

From the above lemma by summing over all edges of a path in $G$ we also get that $Y$ is a $(1+\epsilon)$ spanner for the disk graph $G$. For ball graphs in $\mathbb{R}^k$, $Y$ has $O(n\epsilon^{-k+1})$ edges [Luk99]. For unit ball graphs the modified Yao-graph is same as the original Yao-graph with long edges removed.

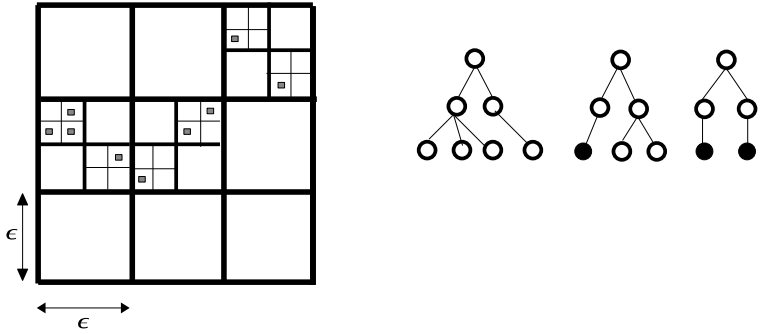

Figure 1: The quad-dissection procedure for a set of points and the corresponding forest $\Gamma$. The donut shaped nodes are the interesting nodes in $\Gamma$.

In the remaining paper we improve upon this result in the following manner. All constructions are for $(1+\epsilon)$-spanner.

**Unit Ball Graphs:** For unit ball graphs in $\mathbb{R}^k$ we solve bichromatic closest pair problems to produce spanners with $O(n\epsilon^{-k+1})$ edges. A spanner can be constructed in time $O(n \log n\epsilon^{-1} + n\epsilon^{-2})$ for $k = 2$ by solving post-office problems. In higher dimensions using the currently best algorithm of Aggarwal *et al.* [AESW91] for solving the bichromatic closest pair problem, we can construct a spanner in expected running time $\tilde{O}(n^{4/3}\epsilon^{-2} + n\epsilon^{-3})$ for $k = 3$ and $O(n^{2-2/(\lceil k/2\rceil+1)+\delta}\epsilon^{-k+1} + n\epsilon^{-k})$ for $k \geq 4$, where $\delta$ is any positive constant. In addition, we show that constructing $(1+\epsilon)$-spanners for unit ball graphs has almost equivalent randomized complexity to solving bichromatic closest pair problem for $n$ points. Additionally the spanners for unit ball graphs have small geometric separator (hereditary), which we exploit to obtain faster approximation algorithms for proximity problems like diameter and distance queries.

**Ball Graphs:** We generalize the above result to ball graphs in $\mathbb{R}^k$. In this case we replace bichromatic closest pair problems by halfspace range searching problems. The spanner has $O(n\epsilon^{-k+1})$ edges. For the interesting case when $\rho(\mathcal{D})$ is polynomially bounded by using the currently best algorithm of Aggarwal and Matoušek [AM95] for half space range searching, we obtain a running time $O(n^{2d+\delta}\epsilon^{-kd} + n\epsilon^{-k})$, where $d = 1 - 1/(\lfloor k/2 \rfloor + 2)$ and $\delta$ is any positive constant.

## 3 Spanners for Unit Disk Graphs

To illustrate our key ideas we first describe an algorithm that produces a spanner $G'$ of unit disk graph $G$ with $O(n\epsilon^{-2})$ edges. Later we describe the procedure for sparsifying the spanner. We first describe a high level idea of our algorithm and prove the claimed stretch factor, and then define an algorithmic version of the construction. The spanner $G'$ is constructed in the following manner.

**Quad-Dissection:** The input to our algorithms is a set of points $(\mathcal{P})$ in $\mathbb{R}^2$. The bounding box of $\mathcal{P}$ is a smallest rectangle enclosing the points and we assume the left bottom corner as the origin. An L-grid is defined by horizontal and vertical line segments drawn at $y = L\mathbb{Z}$ and $x = L\mathbb{Z}$ within the bounding box. Our spanner construction involves recursively partitioning the plane using a simple variant of quadtrees. A quad-dissection of the the L-grid is a recursive partition into smaller squares. We view it as a 4-ary forest with the root nodes as the non-empty squares in the L-grid. Each square is partitioned into four equal squares, which form its children. We continue partitioning the non-empty squares until each disk center is contained in a separate square.

**Constructing the Forest:** Let $\Gamma$ denote the forest from the quad-dissection of the $\epsilon$-grid. Roots represent the non-empty squares of the $\epsilon$-grid. $\Gamma$ is a collection of disjoint trees, each of which is rooted at a node of Roots. Note that the set of nodes at depth $l$ in the forest corresponds to the set of non-empty squares defined by the $\epsilon 2^{-l}$-grid. Define $C(t)$ as set of disk centers lying contained by the node $t$. A node $t$ of the forest is called *interesting* if the set of disk centers contained it is not equal to the set of disk centers in its parent, i.e., $C(t) \neq C(P(t))$. By definition all nodes in Roots are interesting. It immediately follows that $\Gamma$ has at most $2n$ interesting nodes.

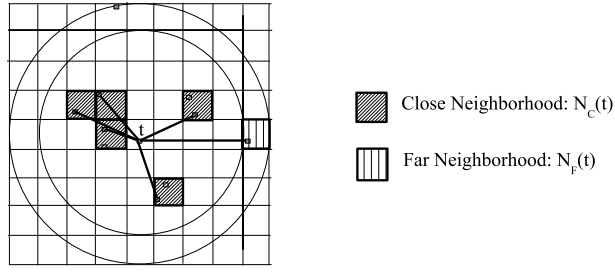

Figure 2: Close and Far neighborhoods of a node t in Roots. Concentric circles are $O(\epsilon)$ apart.

**Choosing the Representatives:** For every leaf node $t \in \Gamma$ we chose the disk center in $C(t)$ as its representative $R_t$. For every internal node $t \in \Gamma$ we pick one of its children $t'$ and set $R_t$ to $R_{t'}$.

**Neighborhood of Nodes:** For every $t \in$ Roots, define its *close neighborhood* $N_C(t)$, as the set of all nodes in Roots which are within a distance of $2 - \sqrt{2}\epsilon$. The distance between squares can be thought as distance between their centers. For every $t \in$ Roots, define its *far neighborhood* $N_F(t)$, as the set of all $t' \in$ Roots which are within distance $(2 - \sqrt{2}\epsilon, 2 + \sqrt{2}\epsilon]$ and for which there exists at least one edge between vertices of $C(t)$ and $C(t')$ in $G$. For every interesting node $t \notin$ Roots, define its close neighborhood $N_C(t)$, as the set of all nodes at depth $d_t$ which are within a distance of $2^{-d_t}$ from $t$. For every interesting node $t \notin$ Roots set its far neighborhood $N_F(t)$ to $\emptyset$.

**Edges in Spanner $G'$:** For every interesting node $t$, we add a edge between $R_t$ and $R_{t'}$, where $t' \in N_C(t)$. Additionally for every node $t \in$ Roots, we put an edge in $G$ between disks $C(t)$ and $C(t')$ into the spanner, where $t' \in N_F(t)$.

**Lemma 2.** *The spanner $G'$ has $O(n\epsilon^{-2})$ edges and is a subgraph of the unit disk graph $G$.*

**Proof.** The nodes at depth $d_t$ in $\Gamma$ are defined by the set of non-empty squares of $\epsilon 2^{-d_t}$-grid. Thus the set $N_C(t)$ is of size $O(\epsilon^{-2})$ and for every interesting node we add $O(\epsilon^{-2})$ edges from its representative $R_t$. There is an edge $(R_t, R_{t'})$ with $t' \in N_C(t)$ in $G$ because the Euclidean distance between them is at most 2.

Also for every $t \in$ Roots, the nodes in $N_F(t)$ are the nodes that lie between two concentric circles $2\sqrt{2}\epsilon$ apart. Since nodes in Roots are defined by the $\epsilon$-grid, there are only $O(\epsilon^{-1})$ nodes in $N_F(t)$. Thus we only add $O(\epsilon^{-1})$ additional edges for every node in Roots and these edges are added only if they are present in $G$. Since there are at most $O(n)$ interesting nodes, we get the claimed result. ❏

**Lemma 3.** *Let $u$ and $v$ be disk centers, such that there exists a node $t$ in $\Gamma$, with $u, v \in C(t)$. Then there exists a path in the spanner $G'$ such that $d_{G'}(u, v) \leq (1 + c_u \epsilon) d(u, v)$ for some constant $c_u$.*

**Proof.** The maximum square length is $\epsilon$. This implies that if there exists a node $t$ with $u, v \in C(t)$, then $d_G(u, v) = d(u, v)$. The proof is by induction over the size of $C(t)$. The base case of $|C(t)| = 1$ is trivial. Now in the subtree rooted at $t$, consider the maximum depth such that there exist nodes $a$ and $b$ with $u \in C(a)$, $v \in C(b)$ with an edge between $R_a$ and $R_b$ in $G'$. There exists such $a$ and $b$ because interesting nodes with the same parent have an edge between their representatives. The case where $u = R_a$ or $v = R_b$ is simpler and for now we assume $u \neq R_a$ and $v \neq R_b$.

In the subtree rooted at $a$, find the closest descendant $a'$ of $a$ which is interesting with $u \in C(a')$. Let $b'$ be the node at same depth as $a'$ with $v \in C(b')$. There exists no edge between between representatives of $a'$ and $b'$ (because $a$ and $b$ are the deepest nodes with this property). This implies that $d(u,v) \geq 2^{-d_{a'}}$. Also if not $a'$, at least $P(a')$ contains both $u$ and $R_a$. Therefore for some constant $c_s$, we get $d(u, R_a) \leq c_s \epsilon 2^{-d_{a'}}$ and thus $d(u, R_a) \leq c_s \epsilon d(u, v)$. Using similar arguments one can show that $d(R_b, v) \leq c_s \epsilon d(u, v)$.

By the inductive hypothesis we know, $d_{G'}(u, R_a) \leq (1 + c_u \epsilon)d(u, R_a)$ and $d_{G'}(R_b, v) \leq (1 + c_u \epsilon)d(R_b, v)$. In $G'$ there exists a path from $u$ to $v$ of length $(1 + c_u \epsilon)d(u, R_a) + d(R_a, R_b) + (1 + c_u \epsilon)d(R_b, v)$. Finally putting together everything we get

$$\begin{aligned} d_{G'}(u,v) &= (1 + c_u\epsilon)d(u, R_a) + d(R_a, R_b) + (1 + c_u\epsilon)d(R_b, v) \\ &\leq (2 + c_u\epsilon)d(u, R_a) + d(u, v) + (2 + c_u\epsilon)d(R_b, v) \leq (1 + c_u\epsilon)d(u, v) \end{aligned}$$

The second inequality comes from applying the triangle inequality and the final step involves substituting $d(u, R_a)$ and $d(R_b, v)$ in terms of $d(u, v)$. The constants satisfy $c_s \leq 2\sqrt{2}$ and $c_u \geq 4c_s/(1 - 2c_s\epsilon)$. ❑

**Lemma 4.** *Let $(u, v)$ be an edge in the unit disk graph $G$. Then there exists a path in the spanner $G'$ such that $d_{G'}(u, v) \leq (1 + c_u \epsilon)d(u, v)$ for some constant $c_u$.*

**Sketch of Proof.** If there exists a node $t$ such that $u, v \in C(t)$, then by Lemma 3 the claim is true. Otherwise, consider the deepest nodes $a, b$ in $\Gamma$, such that (i) $u \in C(a)$, $v \in C(b)$, and (ii) edge between $u' \in C(a)$ and $v' \in C(b)$ is in $G'$. Such $a$ and $b$ exist at least as nodes in Roots. As in Lemma 3 one can argue that $d(u, u') \leq c_s \epsilon d(u, v)$ and $d(v', v) \leq c_s \epsilon d(u, v)$. From Lemma 3, we also know that $d_{G'}(u, u') \leq (1 + c_u \epsilon)d(u, u')$ and $d_{G'}(v', v) \leq (1 + c_u \epsilon)d(v', v)$. The proof can be completed by applying the triangle inequality. ❑

**Sparsifying the Spanner:** Let $\mathcal{C}(t)$ be the collection of $\epsilon^{-1}$ cones such that (i) each cone has apex at $R_t$, (ii) each cone has an opening angle of $2\pi\epsilon$, and (iii) union of these cones covers the entire area represented by $N_C(t)$.

Instead of adding edges to every $R_{t'}$, $t' \in N_C(t)$, we include the shortest edge between $R_t$ and $R_{t'}$ in each cone. Since there are only $\epsilon^{-1}$ cones, we conclude that the number of edges in the spanner is $O(n\epsilon^{-1})$. Using the analysis of Lemma 1 along with Lemma 4, it follows that the graph is a $(1 + \epsilon)$-spanner. Let $G'$ denote this sparsified spanner.

### 3.1 Algorithmic Version

We now describe the algorithmic counterparts for each step in the construction of $G'$.

**Constructing the Forest:** We construct a compressed forest (compressed quadtrees) $\Gamma'$, in which we only introduce the interesting nodes of $\Gamma$ and shortcut the degree one internal nodes. For the sake of construction we introduce some terminology. Given two binary numbers $X = \sum x_i 2^i$ and $Y = \sum y_i 2^i$, define a bitwise shuffle operation of $X$ and $Y$ as:

$$X|Y = \sum_i (2x_i + y_i)4^i$$

Define $\text{agree}(X, Y)$ as the smallest $l$ such that for $X \neq Y$ we have $\lfloor X2^{2l+1} \rfloor \neq \lfloor Y2^{2l+1} \rfloor$. Note that if $X|Y < X'|Y'$, then either $\text{agree}(X, X') \leq \text{agree}(Y, Y')$ and $X < X'$ or $\text{agree}(X, X') > \text{agree}(Y, Y')$ and $Y < Y'$. We also define $\text{agree}(X, Y, l')$, with $X \neq Y$ as the smallest $l$ satisfying $\lfloor X2^{2l+l'+1} \rfloor \neq \lfloor Y2^{2l+l'+1} \rfloor$. Define $\text{round}(X, l) = \lfloor X2^l \rfloor / 2^l$.

For a point $p$, let $x(p)$ and $y(p)$ denote its $x$ and $y$-coordinates and let $p^s = x(p)|y(p)$. Sort the points in $\mathcal{P}$ in increasing order of their shuffled representations. Let $\mathcal{P}_s = \{p_1^s, p_2^s, \ldots, p_n^s\}$ denote this sorted array. Construct a list $A_s$ whose $i^{\text{th}}$ entry $a_i = \text{agree}(p_i^s, p_{i-1}^s, \log \epsilon^{-1})$. Start the forest with a different leaf node for every disk. We maintain a stack with $\text{top}$ representing the last inserted entry and a special bottom of stack entry $(-\infty, \text{Null})$. When we scan a new element $a_i$ in the list $A_s$ we do: (i) if $a_i > a_{\text{top}}$, push $(a_i, t_i)$ onto the top of the stack and increment $i$. (ii) else, pop $(a_{\text{top}}, t_{\text{top}})$ and $(a_{\text{top}-1}, t_{\text{top}-1})$ and create a new node $t$ in $\Gamma'$ with its children as $t_{\text{top}}$ and $t_{\text{top}-1}$. Label edges $(t, t_{\text{top}})$ and $(t, t_{\text{top}-1})$ with $a_{\text{top}}$. Push $(a_{\text{top}-1}, t)$ on the stack.

The depth of a node $t$, $(d_t)$ is the defined by the label of the edge connecting it to its parent. The $\Gamma'$ so generated is the binary tree representation of the compressed quadtree. It can easily be converted into a compressed quadtree by just shortcutting the nodes which have the same depth as their children. The bottom left hand corner of $t$ has coordinates $(\text{round}(x(R_t), d_t + \log \epsilon^{-1}), \text{round}(y(R_t), d_t + \log \epsilon^{-1}))$. $\text{Roots}$ is the set of nodes at depth 0.

The time for constructing the forest is $O(n \log n)$, assuming $\text{agree}^1$ operation can be implemented in constant time. The shuffle operation needn't be performed explicitly as $\text{agree}$ can be used to generate $\mathcal{P}_s$.

**Finding the close neighborhood:** Finding the close neighborhood for nodes in $\text{Roots}$ is straight forward. Because we work with $\Gamma'$ we construct a *close pseudo-neighborhood* $(N_C'(t))$ for a node $t$. The close pseudo-neighborhood consists of a set of nodes, where each element $r$ of the set either belongs to $N_C(t)$ or in $\Gamma$ the node $r$ is the maximal depth ancestor of a node in $N_C(t)$.

Now assuming we have found $N_C'(t)$, we describe a method of constructing $N_C'(t')$ for a child $t'$ of $t$ in $\Gamma'$. We access all the nodes in $N_C'(t)$. For every $a \in N_C'(t)$, we look at the subtree $\Gamma_a'$ rooted at $a$. In this subtree we do a BFS traversal with a modification that the subtree for any $b \in \Gamma_a'$ is accessed only if $d_b \leq d_{t'}$ and $R_b$ is at most distance $(1 + \sqrt{2}\epsilon)2^{-d_b}$ away from $R_{t'}$. The nodes at which the traversal ends and whose representatives are at most $(1 + \sqrt{2}\epsilon)2^{-d_{t'}}$ away from $R_{t'}$ define $N_C'(t')$.

Any node $g \in \Gamma'$ accessed during the traversal has a node $f \in \Gamma$ such that: (i) $f$ is an ancestor of $t'$ in $\Gamma$, (ii) $d_g = d_f$, and (iii) the distance between $f$ and $g$ is $O(2^{-d_g})$. We charge the access to node $g$ to node $f$ ($f$ needn't be in $\Gamma'$). For the entire algorithm we can charge all the accesses to node $g$ to different nodes which are at the same depth as $g$ and only $O(2^{-d_g})$ away (note there are only $O(\epsilon^{-2})$ such nodes). This implies that the time for finding close pseudo-neighborhoods for all nodes is $O(n\epsilon^{-2})$. Alternatively, one can add edges to only the closest representative in each cone within the same time bound.

**Finding the far neighborhood:** The far neighborhood of a node $t \in \text{Roots}$ is computed by first determining the set of nodes $F(t) \subseteq \text{Roots}$ which are at distance $(2 - \sqrt{2}\epsilon, 2 + \sqrt{2}\epsilon]$ away from $t$. For each $t' \in F(t)$, we solve the bichromatic closest pair problem with one set as $C(t)$ and other as $C(t')$. If the disks corresponding to the closest pair intersect, we add the corresponding edge into the spanner. It is well known that bichromatic closest pair in $\mathbb{R}^2$ can be solved in $O(n \log n)$ by using post-office problems. Since each set $C(t)$ takes part in only $O(\epsilon^{-1})$ bichromatic closest pair problems, we get a time of $O(n \log n \epsilon^{-1})$ for finding all far neighborhoods.

We can summarize these results as follows.

---

[1]The authors believe that the same running time can be obtained by just using floor of binary logarithm (a simple operation). Also the floor operation used here is not dangerous as we don't multiply variables.

**Theorem 1.** *There exists an algorithm for producing a $(1+\epsilon)$-spanner for unit disk graphs. The spanner has $O(n\epsilon^{-1})$ edges and can be computed in time $O(n\log n\epsilon^{-1} + n\epsilon^{-2})$.*

## 3.2 Extension to Unit Ball graphs

The quadtrees becomes octrees in $\mathbb{R}^k$. The close neighborhood of any node is of size $O(\epsilon^{-k})$, but again one can sparsify to get it to $O(\epsilon^{-k+1})$. The far neighborhood of any node in Roots is $O(\epsilon^{-k+1})$. Therefore the total number of edges is $O(n\epsilon^{-k+1})$. The bottleneck in the running time comes from solving bichromatic closest pair problems for finding far neighborhoods of nodes in Roots. We now state a general result describing the upper bounds of the spanner construction in terms of the upper bounds for bichromatic closest pair problem.

Let $f(n)$ be any function satisfying $\log f(n) = o(\log n)$, i.e., $f$ grows slower than polynomial. We assume in the rest of discussion that $n^c f(n)$ is an upper bound of on the time for computing a bichromatic closest pair for a total on $n$ points, for some constant $c \geq 1$.

**Lemma 5.** *For unit ball graphs in $\mathbb{R}^k$, the worst case running time for finding the far neighborhood for all the nodes in Roots is $O(n^c f(n)\epsilon^{-k+1})$.*

**Proof.** For every node $t$ in Roots, we solve bichromatic closest pair problems between $C(t)$ and $C(t')$, $t' \in$ Roots and $|C(t')| \leq |C(t)|$. Since $t$ participates in $O(\epsilon^{-k+1})$ bichromatic closest pair problems, the cost of solving these problems (charged to $t$) is $O(|C(t)|^{2c} f(|C(t)|)\epsilon^{-k+1})$. Since every disk center is contained in exactly one node of the Roots, the total cost is

$$O(\sum_{t\in\text{Roots}} |C(t)|^c f(|C(t)|)]\epsilon^{-k+1}) \leq O(n^c f(n)\epsilon^{-k+1})$$

☐

The currently best algorithm of Aggarwal and Matoušek [AESW91] for finding bichromatic closest pair between $P$ red and $Q$ blue points in $\mathbb{R}^k$ runs in expected time $O((PQ\log P\log Q)^{2/3} + P\log^2 Q + Q\log^2 P)$ for $k = 3$ and in expected time $O((PQ)^{1-1/(\lceil k/2\rceil+1)+\delta} + P\log Q + Q\log P)$ for $k \geq 4$, where $\delta$ is any positive constant. We can summarize the results for spanners in unit ball graphs as

**Theorem 2.** *There exists an algorithm for producing a $(1+\epsilon)$-spanner for unit ball graphs in $\mathbb{R}^k$ ($k \geq 3$). The spanner has $O(n\epsilon^{-k+1})$ edges and can be computed in time $O(n^c f(n)\epsilon^{-k+1} + n\epsilon^{-k})$.*

**Corollary 1.** *The spanner $G'$ can be constructed in expected time $\tilde{O}(n^{4/3}\epsilon^{-2} + n\epsilon^{-3})$ for $k = 3$ and for $k \geq 4$ it can be constructed in expected time $O(n^{2-2/(\lceil k/2\rceil+1)+\delta}\epsilon^{-k+1} + n\epsilon^{-k})$, where $\delta$ is any positive constant.*

The above result shows that finding spanner of unit ball graphs in $\mathbb{R}^k$ is not much harder than computing bichromatic closest pair for $n$ points in $\mathbb{R}^k$. In the other direction from the results of Eppstein [Epp04] and Chan [Cha99] we know that the randomized expected time bounds for constructing even the spanning forest of unit balls in $\mathbb{R}^k$ and a Euclidean minimum spanning tree (or a bichromatic closest pair) in $\mathbb{R}^k$ are within constant factors. To make the relation between the running time for spanner construction and bichromatic closest pair problem more precise we introduce a few definitions. Define *exponent* of a problem $\mathcal{A}$ with respect to input size $n$ to be

$$\inf\{c \mid \text{there exists an algorithm for solving } \mathcal{A} \text{ with running time } O(n^c)\}$$

Since the bichromatic closest pair and problem and Euclidean minimum spanning tree have asymptotically same complexities [AESW91, KLN99], their exponents are also equal.

**Theorem 3.** *The exponent of $(1+\epsilon)$-spanner for $n$ unit balls in $\mathbb{R}^k$ is same as the exponent of bichromatic closest pair for $n$ points in $\mathbb{R}^k$.*

**Proof.** From Theorem 2, we know that exponent of the $(1+\epsilon)$-spanner is not more than the exponent of the bichromatic closest pair. For the other direction, once we have the spanner any graph traversal can be used to construct the spanning forest of unit balls. Therefore, from results of Eppstein [Epp04] about randomized equivalence between spanning forest and bichromatic closest pair, we also get that the exponent of bichromatic closest pair is not more than the exponent of $(1+\epsilon)$-spanner. ❑

## 4 Separators in Spanner Graph

In this section we show that the spanner $G'$ has a small geometric vertex separator. Our results can be seen as analogous to results by Smith and Wormald [SW98] for separators in spanners for complete Euclidean graphs. We show that $G'$ has an $n^{1-1/k}\epsilon^{-k+1/2}$-vertex separator, whose removal leaves two sets with each having at most 2/3 of the original vertices (2/3-split). Furthermore, this separator can be found in time $O(n \log n)$. Due to lack of space the entire algorithm and its analysis is presented in Appendix A. We can summarize these results as follows.

**Theorem 4.** *A $n^{1-1/k}\epsilon^{-k+1/2}$-separator decomposition with 2/3-splits of spanner $G'$, for unit ball graphs in $\mathbb{R}^k$ can be found in time $O(n \log n)$.*

### 4.1 Proximity Problems on Spanner

Gao and Zhang [GZ05] discuss many approximate proximity problems for the case of points in plane. Using a well separated pair decomposition for unit disk graph on $n$ points in the plane, they show how to answer $(1+\epsilon)$ approximate distance queries in time $O(1)$ after $O(n\sqrt{n \log n}\epsilon^{-3})$ time preprocessing. They also estimate the diameter of the point set in time $O(n\sqrt{n \log n}\epsilon^{-2})$.

We now note that other than the standard advantages of using a sparse spanner for solving approximate proximity problems, the separator decomposition of the spanner buys us a little more. We briefly describe these advantages for ball graph $G$ in $\mathbb{R}^k$ with $f(n,\epsilon) = n^{1-1/k}\epsilon^{-k+1/2}$.

- **$3/2 + \epsilon$-Diameter Approximation:** Due to lack of space, the algorithm is presented in Appendix B. The running time is of $\tilde{O}(nf(n,\epsilon) + f^3(n,\epsilon))$.

- **Distance Queries:** It is easy to extend the results of shortest distance algorithms for planar graphs from Arikati *et al.* [ACC$^+$96] to spanner $G'$. By using the same technique, after $\tilde{O}(nf(n,\epsilon))$ time and space preprocessing, we can answer an $(1+\epsilon)$-stretch query in $O(f(n,\epsilon))$ and a $(2+\epsilon)$-stretch query in time $O(\log f(n,\epsilon))$.

## 5 Spanners for Disk Graphs

In this section we extend the spanner results for arbitrary disk graphs. Each disk is associated with a level, a disk $D_p$ is of level $l$ if: $2^{-l} \leq r_p < 2^{-l+1}$. Again we first present a high-level description of the algorithm before describing the actual implementation. We now stress the changes in construction for disk graphs from the case of unit disk graphs. Let $G$ denote the disk graph and $G'$ be the constructed spanner.

**Constructing the Forest:** Let $\Gamma$ denote the forest from the quad-dissection of the $\epsilon$-grid. We introduce a disk of level $l$ only at depth $l$. That is for a node $t \in \Gamma$ of depth $d_t$, the set $C(t)$ consists of only disks which are of at most level $d_t$ and their centers lie are contained in node $t$. The definition of an interesting node remains the same.

**Choosing the Representatives:** For leaf nodes representatives are chosen in the same manner. For an internal node $t$, pick a child $t'$ with $R_{t'} \in C(t)$ and set $R_t$ to $R_{t'}$.

**Neighborhood of Nodes:** A node $t$ belongs to Roots if: (i) $d_t = 0$ and $C(t) \neq \emptyset$, i.e., square of $\epsilon$-grid with non-empty $C(t)$, or (ii) $C(t) \neq \emptyset$ and $C(P(t)) = \emptyset$. For every interesting node $t$, define $N_C(t)$ as the set of all nodes at depth $d_t$ which are within a distance of $2^{-d_t}$ from $t$. For every interesting node $t \notin$ Roots, set its far neighborhood $N_F(t)$ to $\emptyset$.

For a node $t \in \Gamma$ and integers $\alpha, \beta$ define its $[\alpha, \beta]$-shift, $(S([\alpha, \beta], t))$ with $\alpha \in [-2\epsilon^{-1}, -(2\epsilon)^{-1}] \cup [(2\epsilon)^{-1}, 2\epsilon^{-1}]$ and $\beta \in [-2\epsilon^{-1}, -(2\epsilon)^{-1}] \cup [(2\epsilon)^{-1}, 2\epsilon^{-1}]$ as the square $t'$ obtained by shifting the x- and y- coordinates of all points in $t$ by $\epsilon \alpha 2^{-d_t}$ and $\epsilon \beta 2^{-d_t}$ respectively. For every $t \in$ Roots, construct $O(\epsilon^{-2})$ ordered buckets, where $\mathsf{bucket}([\alpha, \beta], t)$ is defined as

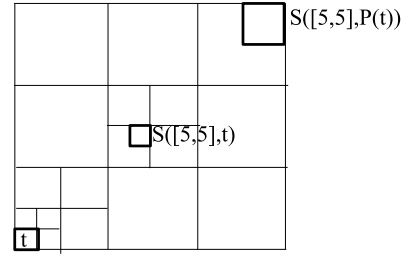

$$\mathsf{bucket}([\alpha, \beta], t) = \{S([\alpha, \beta], t), S([\alpha, \beta], P(t)), S([\alpha, \beta], P(P(t))), \ldots\}$$

Empty squares can be ignored. All nodes in a bucket are also actual nodes in $\Gamma$. For $t \in$ Roots, do the following: scan through every $\mathsf{bucket}([\alpha, \beta], t)$ and find the first node $t' \in \mathsf{bucket}([\alpha, \beta], t)$ such that there exists $v \in C(t)$ and $u \in C(t')$ with edge $(u, v)$ in $G$ Add $t'$ to $N_F(t)$.

**Edges in Spanner $G'$:** Added exactly the same manner as with unit disk graphs.

**Lemma 6.** *The spanner $G'$ has $O(n\epsilon^{-2})$ edges and is a subgraph of disk graph $G$.*

**Proof.** Same as Lemma 2. □

The idea behind creating the buckets is (i) to ensure that there exists lines passing through all the nodes in a single $\mathsf{bucket}$ and $t$, and (ii) to ensure that disks that intersect any disk centered in $t$, have their centers either close or within a node of some $\mathsf{bucket}$. The first fact easily follows from the construction. Before proving the second fact it is convenient to define for a node $t \in \Gamma$,

$$\mathsf{Bucket}(t) = \bigcup_{\alpha, \beta} \mathsf{bucket}([\alpha, \beta], t)$$

with $\alpha, \beta$ in the interval $[-2\epsilon^{-1}, -(2\epsilon)^{-1}] \cup [(2\epsilon)^{-1}, 2\epsilon^{-1}]$.

**Lemma 7.** *Let $u$ be a disk center contained in node $t$. Then for every edge $(u, v)$ in $G$, $v$ is contained in $\mathsf{Bucket}(t) \cup N_C(t)$.*

**Proof.** If $d(u, v) \leq 2^{-d_t}$ then $v$ is contained in $N_C(t)$. If $2^{-d_t} < d(u, v) \leq 2^{-d_t+1}$ then $v$ is contained in $\bigcup_{\alpha, \beta} S([\alpha, \beta], t)$. If $2^{-d_t+1} < d(u, v) \leq 2^{-d_t+2}$ then $v$ is contained in $\bigcup_{\alpha, \beta} S([\alpha, \beta], P(t))$. Similarly for increase in the distance by a factor of 2, we need to move one position in all the buckets of $t$ to finish the proof. □

**Lemma 8.** *Let $(u, v)$ be a edge in the disk graph $G$. Then there exists a path in the spanner $G'$ such that $d_{G'}(u, v) \leq (1 + c_d \epsilon) d(u, v)$ for some constant $c_d$.*

**Proof.** See Appendix C. □

## 5.1 Algorithmic Version

**Constructing the Forest:** We first construct the forest $\Gamma'$ exactly as described for unit disk graphs. At every node we also maintain the disk of the largest radius in it. The representative $R_t$ of $t \in \Gamma'$ is chosen to be the disk of the largest radius, which was also the representative in one of the children of $t$. Then we visit each node $t$ and remove the edge to its parent $P(t)$ if the largest radius among disks with centers in $t$ is less than $2^{-d_t}$. If the edge is removed then $t$ is added to Roots.

**Finding the neighborhoods:** The close pseudo-neighborhood of a node can be obtained as with unit disk graphs. For determining the far neighborhood of a node $t \in$ Roots we make the following distinction.

**Small Global Stretch:** In this we use the observation from Gupta *et al.* [GJS99] that maps the problem of reporting the intersection of a collection of disks with a query disk into halfspace range searching in two dimensions higher. Aggarwal and Matoušek [AM95] construct a data structure for $\mathbb{R}^k$, which for any parameter $n \leq m \leq n^{\lfloor k/2 \rfloor}$ and any positive constant $\delta$, after $O(m^{1+\delta})$ space and time preprocessing answers half-space queries in time $\tilde{O}(n/m^{1/\lfloor k/2 \rfloor})$, and has $O(m^{1+\delta}/n)$ amortized update time. We construct this data structure bottom-up. Let $t$ be a node in $\Gamma'$ with children's $t_i, i = i, \ldots, 4$ satisfying $|C(t_i)| \leq |C(t_4)|$. We construct the data structure for $t$ by inserting $C(t_1)$, $C(t_2)$ and $C(t_3)$ into $C(t_4)$. We then query the data structure with every disk in $C(t')$ with $t \in$ Bucket$(t')$ to check for intersection.

If $\rho(\mathcal{D})$ is the global stretch, then each bucket$([\alpha, \beta], t)$ is of size $O(\log(\rho(\mathcal{D})))$. Therefore $|\text{Bucket}(t)| = O(\log(\rho(\mathcal{D})\epsilon^{-2}))$ and Bucket$(t)$ can be constructed within the same time bound. Each disk acts as a query disk $\log(\rho(\mathcal{D}))\epsilon^{-2}$ times, so total number of queries is at most $O(n \log(\rho(\mathcal{D}))\epsilon^{-2})$ times.

To balance the total time for setting up the data structure at every node and the total query time, we assume the parameter $m$ of [AM95] to be $n^c$ for some $c \geq 1$. As we work in two dimensions higher, each query can be answered in time $\tilde{O}(n^{1-c/2})$. The total time for answering queries is $\tilde{O}(n^{2-c/2} \log(\rho(\mathcal{D}))\epsilon^{-2})$. The total time for setting up all the data structures of [AM95] by the procedure described above is $O(n^{c+\delta})$, where $\delta$ is any positive constant. On eliminating the constant $c$, by balancing the exponents of $n$ in these two running times, we get a running time of $O(n^{4/3+\delta} \log^{2/3}(\rho(\mathcal{D}))\epsilon^{-4/3} + n\epsilon^{-2})$ for finding the far neighborhood of all nodes. The space used for storing the data structures is $O(n^{c+\delta})$.

**Large Global Stretch:** In this case we use adjacency lists to place the edges $(u, v)$ satisfying (i) $u \in C(t)$, $v \notin C(t)$, and (ii) $r_u \leq r_v$ in the right bucket. Consider the line segment (edge) connecting $u$ and $v$, and find points $v'$ and $v''$ on the line segments such that $2d(u, v') = d(u, v'')$ and $v'$ is contained in $\bigcup_{\alpha, \beta} S([\alpha, \beta], t)$ and $v''$ is not contained in $\bigcup_{\alpha, \beta} S([\alpha, \beta], t)$ with the usual ranges of $\alpha$ and $\beta$. Let $\alpha', \beta'$ be such that $v'$ is contained in $S([\alpha', \beta'], t)$. We put edge $(u, v)$ in bucket$([\alpha', \beta'], t)$. Once all the edges have been put into their respective bucket, we pick the shortest edge in each bucket and add it to the spanner. The entire procedure can be implemented in $O(|E|)$ given the adjacency list.

## 5.2 Extension to Ball Graphs

The extension to higher dimension can be carried out as described for unit ball graphs. For balls graphs in $\mathbb{R}^k$, the size of $N_C(t) = O(\epsilon^{-k})$ and $N_F(t) = O(\epsilon^{-k})$. Again we can save a factor of $\epsilon^{-1}$ by sparsifying as for unit ball graphs. In $\mathbb{R}^k$ using results of range-searching from [AM95] we get a running time of $O(n^{2d+\delta} \log^d \rho(\mathcal{D}) \epsilon^{-kd} + n\epsilon^{-k})$, where $d = 1 - 1/(\lfloor k/2 \rfloor + 2)$. For large global stretch the running time is $O(|E| + n\epsilon^{-k})$ as we use adjacency lists. We can summarize the

results as

**Theorem 5.** *There exists an algorithm for producing a $(1+\epsilon)$-spanner for ball graphs in $\mathbb{R}^k$. The spanner has $O(n\epsilon^{-k+1})$ edges and for any positive constant $\delta$ can be constructed in time $O(min\{n^{2d+\delta}\log^d \rho(\mathcal{D})\epsilon^{-kd}, Adj(\mathcal{D})\} + n\epsilon^{-k})$, where $d = 1 - 1/(\lfloor k/2 \rfloor + 2)$ and $Adj(\mathcal{D})$ is the time for constructing the adjacency list for $\mathcal{D}$.*

## 6 Concluding Remarks

A special case of disk graphs is when all disks have almost the same radius (between $[1-\epsilon, 1]$). In that case our spanner construction for unit disk graphs still works by using Sharir's [Sha85] red-blue disk intersection algorithm instead of bichromatic closest pairs. Therefore we get a running time of $O(n\log^2 n\epsilon^{-1} + n\epsilon^{-2})$. Additionally all our results on spanners also extend to cases when intersections are between squares, regular polygons, or congruent axis-parallel rectangles, as well as their higher dimensional versions.

The bottleneck in the running time of our algorithms is the time for finding the intersections between sets of balls. Faster algorithms for finding the intersections would improve the running time of the construction.

# 7  Appendix A: Separators in Spanner Graph for Unit Ball Graphs

Let $G'$ be the spanner for unit ball graph $G$ constructed in Section 3. We now show that the $G'$ has a $\sqrt{n}\epsilon^{-3/2}$-vertex separator and can be found in $O(n \log n)$. Let $\mathcal{T}(l)$ be the set of nodes in $\Gamma'$ at depth $l$. Let $\mathcal{E}(l)$ be the set of all edges that were added to $G'$ when the nodes in $\mathcal{T}(l)$ were considered. Let $\mathcal{D}(l)$ be set of disk centers $u$ with $(u,v) \in \mathcal{E}(l)$.

We use recursive partitions of rectangles. Let $\mathcal{X}$ be the list of disk centers sorted by x-coordinates. Let $\mathcal{X}(\mathcal{R})$ be vertices in rectangle $\mathcal{R}$ sorted by their x-coordinates (similarly we define $\mathcal{Y}$ and $\mathcal{Y}(\mathcal{R})$ for y-coordinates). We say a vertex crosses a line segment if any edge incident on it in $G'$ crosses the line segment. Define a rectangle to be *active* if at least $2n/3$ of the disk centers are inside it and not more than $O(\sqrt{n}\epsilon^{-1})$ vertices inside it crosses the boundary. The initial rectangle is the bounding box (and is active).

A *vertical double line separators* of an active rectangle is a set of at most two vertical line segments that partition the active rectangle, such that only $O(\sqrt{n}\epsilon^{-3/2})$ vertices of $G'$ crosses each line segment (similarly define *horizontal double line separators*). Our algorithm recursively partitions an active rectangle alternatively with a horizontal or a vertical double line separators and stops when there is no more active rectangle.

In order to find the line separators we maintain the following data structure. We partition all the x- coordinates in $\mathcal{X}$ into intervals where every interval has at least $\sqrt{n}\epsilon^{-3/2}$ and at most $3\sqrt{n}\epsilon^{-3/2}$ consecutive points in $\mathcal{X}$. The length of the interval $[x_i, x_j]$ is the difference between the coordinates of its first and last element, i.e., $|x_i - x_j|$. We maintain a balanced binary search tree where leaves represent these intervals and each internal node stores a splitting value between its left and right subtree. Additionally with each node we store the largest length in its subtree. Similarly we construct a tree for $\mathcal{Y}$.

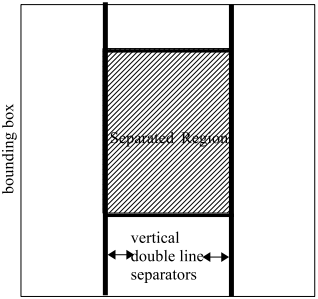

**Constructing double line separators for level l:** Let $\mathcal{R}$ be the currently active rectangle. Let $\mathcal{X}(\mathcal{R}) = \{x_1, x_2, \ldots\}$. We maintain sorted doubly linked lists for both the x- and y- coordinates with pointers between elements representing same vertex point in two lists. Therefore in going to $\mathcal{R}$ from previous active rectangle $\mathcal{R}'$, the lists $\mathcal{X}(\mathcal{R})$ and $\mathcal{Y}(\mathcal{R})$ can be constructed in time proportional to number of points removed from $\mathcal{R}'$ to $\mathcal{R}$.

Let $x_m$ be the median in $\mathcal{X}(\mathcal{R})$. We first find the leaf node containing the median. We then walk up the tree until we find the first left parent having an interval of length greater than $2^{-l+2}$ in its subtree. We access the rightmost leaf in the subtree representing the interval $[x_j, x_{j'}]$ with $|x_j - x_{j'}| \geq 2^{-l+2}$. Define $L_1$ to be a vertical line segment in R to the left of $x_{j'}$ at a distance $2^{-l+1}$. Symmetrically we construct $L_2$ by considering only right parents.

For updating the intervals when we go from $\mathcal{R}'$ to $\mathcal{R}$ we use two operations: *merge* and *split and merge*. When an interval becomes smaller than $\sqrt{n}\epsilon^{-3/2}$ we merge it with its left or right interval whichever is smaller than $2\sqrt{n}\epsilon^{-3/2}$. If neither satisfies the condition we split one of them into two halves and then do the merge. Both of these operations and the subsequent updates to the tree can be performed in $O(\log n)$ time and are only done when an actual vertex is deleted.

$L_1$ and $L_2$ divide $\mathcal{R}$ into at most 3 rectangles, of which at most one forms the new active rectangle. In this new active rectangle $\mathcal{R}''$ we repeat the same procedure with $\mathcal{Y}(\mathcal{R}'')$ to find horizontal line segments $L_3$ and $L_4$. The active rectangle thus formed is associated with the level following l in $\Gamma'$.

**Lemma 9.** *The distance between $L_1$ and $L_2$ at depth l is less than $2^{-l+2}\epsilon^{3/2}\sqrt{n}$.*

**Proof.** Starting from the median we move $L_1$ (or $L_2$) only because there are $\sqrt{n}\epsilon^{-3/2}$ disk centers to the left (or right) within a distance of $2^{-l+2}$. As there are only $n$ disks, the result follows. □

**Remark.** The same distance bound holds also for the separation between $L_3$ and $L_4$. As a consequence of the above lemma we also get that length of $L_1$ or $L_2$ for $l \geq 1$ is at most $2^{-l_p+2}\epsilon^{3/2}\sqrt{n}$ where $l_p$ is the level preceding l in $\Gamma'$. We now show that $L_1$ and $L_2$ form vertical double line separators for $\mathcal{R}$. The arguments for $L_3$ and $L_4$ are similar.

**Lemma 10.** *Let $\mathcal{R}$ be an active rectangle of depth l. $L_1$ and $L_2$ define a vertical double line separators for $\mathcal{R}$.*

**Sketch of Proof.** We work with $L_1$. The case for $L_2$ is symmetric. Since $\mathcal{R}$ is an active rectangle there are only $O(\sqrt{n}\epsilon^{-3/2})$ vertices, having edges of type $(u, v)$, $u \notin \mathcal{X}(\mathcal{R})$ and $v \in \mathcal{X}(\mathcal{R})$ that cross $\mathcal{R}$, and possibly $L_1$. In the remaining cases we only consider edges with both endpoints in $\mathcal{R}$. The edges in $\bigcup_{l' \geq l} E(l')$ have length at most $2^{-l+1}$. Any edge $(u,v) \in \bigcup_{l' \geq l} \mathcal{E}(l')$ crossing $L_1$ must have both u and v be within a distance of $2^{-l+1}$ from $L_1$. This implies that the number of edges in $\bigcup_{l' \geq l} \mathcal{E}(l')$ and hence the vertices in $\bigcup_{l' \geq l} \mathcal{D}(l')$ that cross $L_1$ is $O(\sqrt{n}\epsilon^{-3/2})$. So if $l = 0$ we are done.

We finish the proof by showing that not more than $O(\sqrt{n}\epsilon^{-1})$ vertices in $\bigcup_{l'<l} \mathcal{D}(l')$ cross $L_1$. Fix any level $l' < l$. From Lemma 9, and the remark above we know that the length of $L_1$ is less than $2^{-l_p+2}\sqrt{n}\epsilon^{3/2}$. In $\mathcal{R}$ consider the rectangular region between $L_1$ and a line segment drawn at distance $2^{-l'+1}$ from $L_1$. The area of this region is $2^{-l_p+2}2^{-l'+1}\epsilon^{3/2}\sqrt{n}$ and any edge of $\mathcal{E}(l')$ that crosses $L_1$ will have one endpoint in it. The number of nodes of $\mathcal{T}(l')$ in this region is $2^{l'-l_p+3}\sqrt{n}\epsilon^{-1/2}$. For each node of $\mathcal{T}(l')$ at most $\epsilon^{-1}$ vertices contained in it could be present in $\mathcal{D}(l')$. Thus the number of vertices in $D(l')$ in $\mathcal{R}'$ that cross $L_1$ is at most $2^{l'-l_p+3}\sqrt{n}\epsilon^{-3/2}$. This forms a decreasing geometric series in $l'$. For levels $l''$ which are more than $O(\log n)$ below $l$, only constant number of disks of $\mathcal{D}(l'')$ crosses $L_1$. Our method of picking representatives ensures that they are the same over all such $l''$. Summing over all $l' < l$ we get that the total number of vertices in $\bigcup_{l'<l} \mathcal{D}(l')$ crossing $L_1$ is

$$\sum_{0 \leq l' \leq l_p} 2^{l'-l_p+3}\sqrt{n}\epsilon^{-3/2} = O(\sqrt{n}\epsilon^{-3/2})$$

❑

**Final Shape of Separator:** We stop the algorithm if no rectangles are active. The rectangle with the largest number of disk centers form one component of the separator. The vertices of it which are connected to the outside are the vertex separators.

The proof of termination is omitted in this extended abstract. The algorithm also extends to higher $k$ and we get a $n^{1-1/k}\epsilon^{-k+1/2}$-separator. The rectangle gets replaced by k-dimensional box and line segments are $(k-1)$-dimensional hyperplanes. We recursively partition along every dimension.

## Appendix B: Diameter Approximation

We use a rooted binary tree $T_G$ to represent a separator decomposition of $G = (V, E)$. For a set $V'$ of vertices in $G$, we use $N(V')$ to denote the neighborhood of $V'$. Each node $t \in T_G$ is labeled by two subsets of vertices $V(t) \subseteq V$ and $S(t) \subseteq V(t)$. Let $G(t) = (V(t), E(t))$ denote the subgraph induced by $V(t)$. Then $S(t)$ is the separator in $G(t)$. The root $r \in T_G$ has $V(r) = V$ and $S(r)$ is a separator in $G$. For any $t \in T_G$, the labels of its children $t_0, t_1$ are defined as follows: Let $V_1 \subset V(t)$ and $V_2 \subset V(t)$ be the components separated by $S(t)$ in $G(t)$. Then $V(t_0) = V_1 \cup (S(t) \cap N(V_1))$, $V(t_1) = V_2 \cup (S(t) \cap N(V_2))$.

We describe our 3/2-diameter approximation algorithm for any graph $G = (V, E)$ from the family of weighted directed graphs having a $M^\mu$-separator decomposition, i.e., every subgraph of size $M$ has an $M^\mu$-separator. The algorithm is similar to the algorithm for 3/2-diameter approximation of Aingworth et al. [ACIM99] for general graphs. On graphs with $m$ edges their algorithm runs in time $O(m\sqrt{n\log n} + n^2 \log n)$.

For $e \in E$, let $w(e)$ denote the weight of edge $e$. We associate *boundary vertices*, $B(t)$ with each node $t \in T_G$. The boundary of the root $r \in T_G$ is $B(r) = \emptyset$. The boundary of every other node $t$ is defined as $B(t) = S(P(t)) \cup B(P(t)) \cap V(t)$, where $P(t)$ is the parent of $t$ in $T_G$. We discuss the procedure for single node $t \in T_G$.

**Performing SSSP:** Do Single Source Shortest Path (SSSP) from $S(t)$ on the graph $H(t) = (V(t), E(t) \cup B(t) \times B(t))$, with weights $w_H$ defined as follows: For a edge $e = (v_1, v_2) \in B(t) \times B(t)$ define $w_{H(t)}(e)$ to be $d_{G(t)}(v_1, v_2)$. One can inductively see that these weights are available. For $t = r$ it is true. If $v_1$ or $v_2$ is in $B(P(t))$, inductively we know that weights are available. Otherwise both $v_1$ and $v_2$ are in $S(P(t))$ and we get the weights after doing SSSP on $H(P(t))$ (see Lemma

11). For edge $e \in E(t)$, $w_{H(t)}(e) = w(e)$.

**Estimate for a node:** Let $\max_1$ be the longest distance obtained in $H(t)$ after performing the SSSP from all vertices in $S(t)$. For $u \in V(t_0) \cup V(t_1)$, let $clo(u)$ denote the closest vertex in $S(t)$. Let $z$ be a vertex with $clo(z) = \max\{clo(u)\}$. For $z \in V(t_j)$, do SSSP from $z$ on $H(t)$ to find a farthest node $w \in V(t_{1-j})$, $j = 0, 1$. Let $\max_2$ be the distance between $z$ and $w$. Compute the estimated diameter for node $\max(t)$ as $\max\{\max_1, \max_2\}$.

**Output of Algorithm:** Output the estimated diameter $Dia$ as $\max\{\max(t) : t \in T_G\}$.

**Lemma 11.** *For any $t \in T_G$ and $u, v \in V(t)$, $d_{H(t)}(u, v) = d_G(u, v)$.*

**Proof.** Proof by induction. If $t = r$, the claim is true. By the inductive hypothesis $d_{H(P(t))}(u, v) = d_G(u, v)$. Let $P = \{u, p_1, p_2, \ldots, v\}$ denote the shortest path from $u$ to $v$ in $H(P(t))$. If $P \subseteq V(t)$ then the shortest path is also present in $H(t)$. Otherwise let $i$ be smallest index such that $p_i \in S(P(t))$ and let $j$ be the largest index such that $p_j \in S(P(t))$. Both $p_i$ and $p_j$ are in $B(t)$ and have an edge connecting them with weight $w_{H(t)}(p_i, p_j) = d_G(p_i, p_j)$. Thus distances are preserved in $H(t)$. □

**Lemma 12.** *Let $G$ be a graph with $n$ vertices and an $M^\mu$-separator decomposition, $0 < \mu < 1$, with $\alpha$-splits, then number of edges in $G$ is $O(n)$.*

**Proof.** The number of edges $E(n)$ in $G$ satisfies the recurrence: $E(n) \leq \max_\alpha\{E(\alpha n + Cn^\mu) + E((1-\alpha)n + Cn^\mu)\}$, where $C$ is a given constant from the size of the separator. We can show by induction that $E(n) \leq cn - dn^\mu$ for constants $n_0, c$ and $d$ chosen as follows : (i) choose $n_0$ such that $n - (2Cn^\mu)/(\alpha^\mu + (1-\alpha)^\mu - 1) \geq 1$ for all $n \geq n_0$, (ii) choose $c = \binom{n_1}{2}$, where $n_1 = \max\{n_0/\alpha, n_0/(1-\alpha)\}$, and (iii) choose $d$ such that $2C/(\alpha^\mu + (1-\alpha)^\mu - 1) \leq \frac{d}{c}$. □

**Theorem 6.** *Let $G$ be a positive weighted graph with an $M^\mu$-separator decomposition. The output $Dia$ provides a $3/2$ approximation for the diameter in $G$ and can be computed in time $\tilde{O}(n^{1+\mu} + n^{3\mu})$.*

**Proof.** Let $a$ and $b$ be the two vertices whose shortest path defines the diameter. Let $d_G(a, b) = \triangle$. Let $t$ be the node in $T_G$ with $a \in V(t_0)$ and $b \in V(t_1)$, where $t_0, t_1$ are the children of $t$ in the separator decomposition tree. If either $a \in S(t)$ or $b \in S(t)$ our estimate for diameter equals $\triangle$ (from Lemma 11). We now consider two cases:

**Case 1:** $\forall u \in V(t_0) \cup V(t_1)$, $\exists v \in S(t)$ such that $d_G(u, v) < \triangle/3$: Let $w$ be the node in $S(t)$, s.t. $d_G(a, w) \leq \triangle/3$. Then definitely $d_G(w, b) \geq 2\triangle/3$ . This implies that we get $3/2$ approximation for the diameter as we do SSSP from $w$.

**Case 2:** Otherwise: At least one node exists in $V(t_0) \cup V(t_1)$ to which the distance from all the nodes in $S(t)$ is at least $\triangle/3$. This implies that $z \in V(t_0) \cup V(t_1)$ is also at least $\triangle/3$ from all the nodes in $S(t)$. Assume $z \in V(t_1)$. If the farthest node from $z$ is at a distance $2\triangle/3$, this distance is our estimate and we are done. Otherwise, let $p_1 \in S(t)$ be a node through which the shortest path from $z$ to $a$ passes. $d_G(z, a) = d_G(z, p_1) + d_G(p_1, a) \leq \triangle$. Since $d_G(z, a) \leq 2\triangle/3$ and $d_G(z, p_1) \geq \triangle/3$. This implies $d_G(p_1, a) \leq \triangle/3$. Since $d_G(a, b) = \triangle$, implying $d_G(b, p_1) \geq 2\triangle/3$. Again we get a $3/2$ approximation when we do SSSP from $p_1$.

To analyze the running time, note that for all $t \in T_G$, $|B(t)| = O(n^\mu)$, therefore at every $t$ we introduce at most $O(n^{2\mu})$ edges. The total running time is dominated by the cost of doing SSSP. Using Dijkstra for SSSP and Lemma 12 gives the required bounds. □

# Appendix C: Proof of Lemma 8 from Section 5

**Proof.** Assume w.l.o.g. that $r_u \leq r_v$. Let $a$ be the deepest node in Roots containing point $u$. Let $t$ be the node containing $v$ with $d_a = d_t$. We divide the analysis into parts based on the position of $t$. For the case when $t \in N_C(a)$ the arguments are similar to those in Lemma 4 and one can show $d_{G'}(u,v) \leq (1+c_u\epsilon)d(u,v)$.

If $t \notin N_C(a)$. Consider $\text{bucket}([\alpha',\beta'],a)$ containing $v$. From Lemma 7 we know such $\alpha',\beta'$ exists (as we are in the case where $v$ is not contained in $N_C(t)$). Let $b \in \text{bucket}([\alpha',\beta'],a)$ be a node containing $v$. Let $f \in \text{bucket}([\alpha',\beta'],a) \cap N_F(a)$. Let $(u',v')$ with $u' \in C(a)$ and $v' \in C(f)$ be the edge added to the spanner. The edge $(u,u')$ also exists in $G$ because the the minimum radius disk in $C(a)$ is at least $\epsilon^{-1}$ times $a$'s length. In $G'$ we know from the previous case that $d_{G'}(u,u') \leq (1+c_u\epsilon)d(u,u')$.

Let $a_1$ and $a_2$ be the ancestors of $a$ in $\Gamma$ at depth $d_b$ and $d_f$ respectively. Let $f_1$ be the ancestor of $f$ in $\Gamma$ at depth $d_b$. In the following description the distance between two squares is the distance between their centers and distance between a point and a square is the distance of the point from the center of the square.

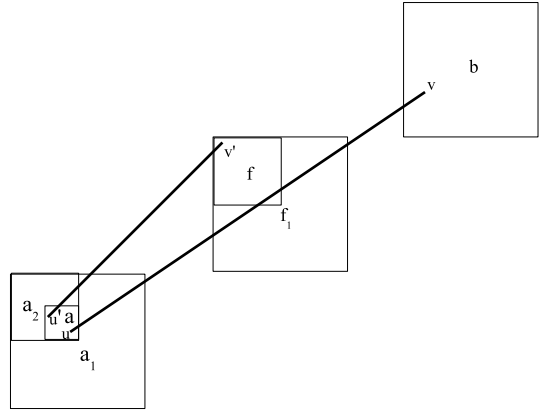

We first claim that $d_{G'}(u,v) \leq (1+c_1\epsilon)d(a_1,b)$ for some constant $c_1$. Since $f$ is obtained by shifting $a_2$ by $\epsilon\alpha'2^{-d_{a_2}}$ and $\epsilon\beta'2^{-d_{a_2}}$, we get $d(u,a_2) \leq c_s\epsilon d(a_2,f)$, $d(f,v') \leq c_s\epsilon d(a_2,f)$, and $d(u,u') \leq c_s\epsilon d(u,v')$. Applying the triangle inequality we obtain

$$d(u,v') \leq d(u,a_2) + d(a_2,f) + d(f,v') \leq (1+2c_s\epsilon)d(a_2,f)$$

In $G'$ we have

$$d_{G'}(u,v') = d_{G'}(u,u') + d(u',v') \leq (1+c_u\epsilon)d(u,u') + d(u',u) + d(u,v') \leq (1+c_2\epsilon)d(u,v')$$

Substituting for $d(u,v')$ implies that $d_{G'}(u,v') \leq (1+c_1\epsilon)d(a_2,f)$ for some constant $c_1$. Note that $d(a_2,f) = d(a_1,f_1)$. To complete the claim we use induction over distances. Since the edge $(v',v)$ also exists in the graph by the inductive hypothesis we get $d_{G'}(v',v) \leq (1+c_1\epsilon)d(f_1,b)$. By construction $d(a_1,b) = d(a_1,f_1) + d(f_1,b)$, therefore we get $d_{G'}(u,v) \leq (1+c_1\epsilon)d(a_1,b)$.

To complete the proof we repeat similar arguments to show that $d(a_1,b) \leq (1+c_u\epsilon)d(u,v)$. Substituting for $d(a_1,b)$ in terms of $d(u,v)$ completes the proof. ☐